\documentclass[conference]{IEEEtran}
\IEEEoverridecommandlockouts
\usepackage{amsmath,amssymb,amsfonts}
\usepackage{algorithmic}
\usepackage{graphicx}
\usepackage{textcomp}
\usepackage{xcolor}
\usepackage{biblatex}
\def\BibTeX{{\rm B\kern-.05em{\sc i\kern-.025em b}\kern-.08em
    T\kern-.1667em\lower.7ex\hbox{E}\kern-.125emX}}
    
\bibliography{conference_101719.bib}
\begin{document}

\title{Commissioning and first measurements of the initial X-ray and $\gamma$-ray detectors at FACET-II\*\\
\thanks{The work at LOA was supported by the ANR (UnRIP project, Grant No. ANR-20-CE30-0030). The work at LOA was also supported by the European Research Council (ERC) under the European Union’s Horizon 2020 Research and Innovation Program (M-PAC project, Grant Agreement No. 715807). Work supported in part by the U.S. Department of Energy under contract number DE-AC02-76SF00515. The work of EG and DAR were supported by the U.S. Department of Energy, Office of Science, Fusion Energy Sciences under Award DE-SC0020076. The work of HE was supported by the Knut and Alice Wallenberg Foundation (KAW 2018.0450). The work of MV and TG is supported by the Portuguese Science Foundation grant FCT No. CEECIND/01906/2018,  CEECIND/04050/2021 and PTDC/FIS- PLA/3800/2021. The UCLA contribution was made possible by DOE grant DE-SC0010064:0011}
}
\author{\IEEEauthorblockN{P.~San~Miguel~Claveria$^{1,2,}$\IEEEauthorrefmark{1}, D.~Storey$^{3}$, G.~J.~Cao$^{4}$, A.~Di~Piazza$^{5}$, H.~Ekerfelt$^{3}$, S.~Gessner$^{3}$, E.~Gerstmayr$^{6}$\\ T.~Grismayer$^{2}$, M.~Hogan$^{3}$, C.~Joshi$^{7}$, C.~H.~Keitel$^{5}$, A.~Knetsch$^{1}$, M.~Litos$^{8}$, A.~Matheron$^{1}$, K.~Marsh$^{7}$, S.~Meuren$^{3}$\\ B.~O'Shea$^{3}$, D.~A.~Reis$^{6}$, M.~Tamburini$^{5}$, M.~Vranic$^{2}$, J.~Wang$^{9}$, V.~Zakharova$^{1}$, C.~Zhang$^{7}$, S.~Corde$^{1}$}
\IEEEauthorblockA{$^{1}$Laboratoire d'Optique Appliquée, ENSTA Paris, CNRS, Ecole Polytechnique, Institut Polytechnique de Paris, Palaiseau, France}
\IEEEauthorblockA{$^{2}$GoLP/Instituto de Plasmas e Fusão Nuclear, Instituto Superior Técnico, Lisboa, Portugal}
\IEEEauthorblockA{$^{3}$SLAC National Accelerator Laboratory, Menlo Park, CA 94025, USA}
\IEEEauthorblockA{$^{4}$Department of Physics, University of Oslo, Oslo, Norway}
\IEEEauthorblockA{$^{5}$Max-Planck-Institut f\"ur Kernphysik, Heidelberg, Germany}
\IEEEauthorblockA{$^{6}$Stanford PULSE Institute, SLAC National Accelerator Laboratory, Menlo Park, CA 94025, USA}
\IEEEauthorblockA{$^{7}$University of California Los Angeles, Los Angeles, CA 90095, USA}
\IEEEauthorblockA{$^{8}$Department of Physics, Center for Integrated Plasma Studies, University of Colorado Boulder, Boulder, Colorado 80309, USA }
\IEEEauthorblockA{$^{9}$Department of Physics and Astronomy, University of Nebraska–Lincoln, Lincoln, NE 68588, USA}
\IEEEauthorrefmark{1}ORCID-IDs: 0000-0001-9342-1573
}

\maketitle

\begin{abstract}
The upgraded Facility for Advanced Accelerator Experimental Tests (FACET-II) at SLAC National Accelerator Laboratory has been designed to deliver ultra-relativistic electron and positron beams with unprecedented parameters, especially in terms of high peak current and low emittance. For most of the foreseen experimental campaigns hosted at this facility, the high energy radiation produced by these beams at the Interaction Point will be a valuable diagnostic to assess the different physical processes under study. This article describes the X-ray and $\gamma$-ray detectors installed for the initial phase of FACET-II. Furthermore, experimental measurements obtained with these detectors during the first commissioning and user runs are presented and discussed, illustrating the working principles and potential applications of these detectors. 

\end{abstract}

\begin{IEEEkeywords}
X-ray detectors, $\gamma$-ray detectors, beam-plasma interaction, Strong-Field QED
\end{IEEEkeywords}

\section{Introduction}
The detection and measurement of high energy photons (X-rays and $\gamma$-rays) produced in laboratory remains an important challenge in several experimental contexts, largely due to the low probability of interaction of these photons with matter. Furthermore, since the cross-section of the photon-matter interactions have a strong dependence on the incoming photon energy, these detectors are typically designed for a defined energy range, outside of which their sensitivity significantly drops. At the new Facility for Advanced Accelerator Experimental Tests (FACET-II), which accelerates electron - and ultimately positron - bunches to maximum delivered beam energy of 13 GeV using the middle kilometer of Linear Accelerator (LINAC) of the SLAC National Accelerator Laboratory, several experimental campaigns rely on the X-ray and $\gamma$-ray photons produced by these particle beams to retrieve valuable information of the key physical processes under study. Yet, depending on the experiment, the relevant spectral range of these photons can span from several to hundreds of keV all the way up to the beam energy.

\begin{figure}
    \centering
    \includegraphics[width=0.48\textwidth]{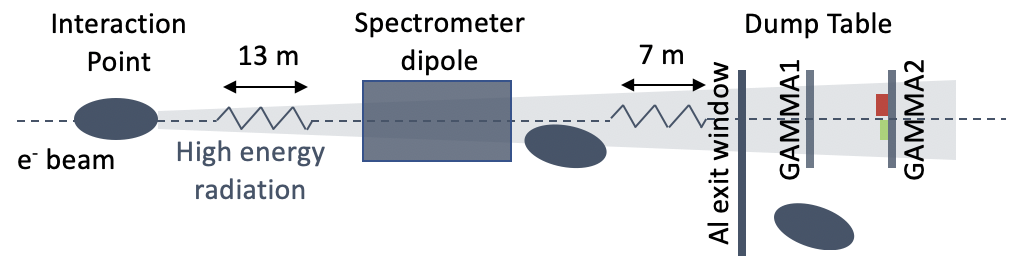}
    \caption{Sketch of the relevant beam line elements for the high energy radiation measurement at FACET-II.}
    \label{fig_intro}
\end{figure}

At FACET-II, the X-ray and $\gamma$-ray photons are produced at the Interaction Point (IP) where the relativistic electrons interact with matter and/or with an Ultra-High-Intensity (UHI) laser pulse. 
After the interaction, these high energy photons co-propagate in vacuum with the relativistic particle bunch until the electrons are deflected down by the imaging spectrometer dipole magnet placed $\approx 13$ m downstream of the IP. Both the photons and electrons exit the vacuum pipe through a 5~mm thick Al window placed $\approx 20$ m downstream of the IP, after which they are detected at the so-called Dump Table, an optical table immediately prior to the beam dump where the detector hardware is mounted (see Fig.~\ref{fig_intro}). At the location of the dump table, the photon axis is vertically separated by $\approx 60$ mm from the dispersed electron axis for the nominal deflection of the spectrometer dipole magnet. This article starts with a brief introduction of the working principles of these high energy radiation detectors together with a description of the installed hardware. Afterwards, the article shows how different experiments at FACET-II benefit from these detectors, presenting the first X-ray and $\gamma$-ray measurements obtained during the initial user-assisted commissioning runs.

\section{Design of GAMMA detectors at FACET-II}

Prior to the initial runs of FACET-II, a collaborative effort involving users from different experiments was carried out with the goal of developing a unified set of diagnostics that would meet the measurement requirements of the initially planned experiments. Based on the outcomes of several simulation tools described in the next paragraphs, the first set of two scintillation-based detectors were developed and installed. Referred to as GAMMA1 and GAMMA2, they are designed to acquire X-ray/gamma-ray angular and spectral information.

When a high energy photon propagates through the scintillator, some of its energy is deposited in the bulk of the material, some of which is then transformed to visible light. The amount of deposited energy depends on the choice of the scintillation material, but also on the incoming photon energy. For instance, a photon with energy $\lesssim$ 1 keV will deposit, on average, almost its entire energy in most commercially available scintillators, i.e. it will be absorbed. In contrast, a 1 GeV photon will deposit a much smaller fraction of its energy in the passage through the scintillator screen. In this article we will refer to the \textit{spectral response} of the detector as the average fraction of the incoming photon energy $\hbar \omega$ that is deposited in the scintillator $\Gamma_{\rm dep}(\hbar \omega)$. Given the wide photon spectra produced during different experimental campaigns at FACET-II, understanding the spectral response of these X-ray and $\gamma$-ray detectors over the relevant photon energies is of key importance to understand the sensitivity of these detectors.

The spectral responses have been calculated using the GEANT4 simulation toolkit~\cite{GEANT4_2003}. The main advantage of using GEANT4 with respect to tabulated data of X-ray absorption is to account for the photon-matter interactions that happen prior to the interaction of the incoming photons with the scintillation screens. Namely, these simulations account for the secondary particle production that can ultimately contribute to the total energy deposited in the scintillator. 
For this purpose, the angular distributions of these secondary particles as well as the spatial distribution of the detectors need to be accounted for in the simulations. 
In the GEANT4 simulations used here, this was achieved by tracking both the primary and secondary particles through a simplified version of the FACET-II geometrical set-up. 
In these simulations all the scintillating screens, as well as other elements on the photon axis, had a transverse spatial extent of $10\times10\;\rm cm^{2}$, corresponding to the acceptance angle of the different vacuum components from the IP to the GAMMA1 and GAMMA2 detectors. Furthermore, all these elements were placed with the same longitudinal spacing as in the experimental set-up.

\begin{figure}
    \centering
    \includegraphics[width=0.48\textwidth]{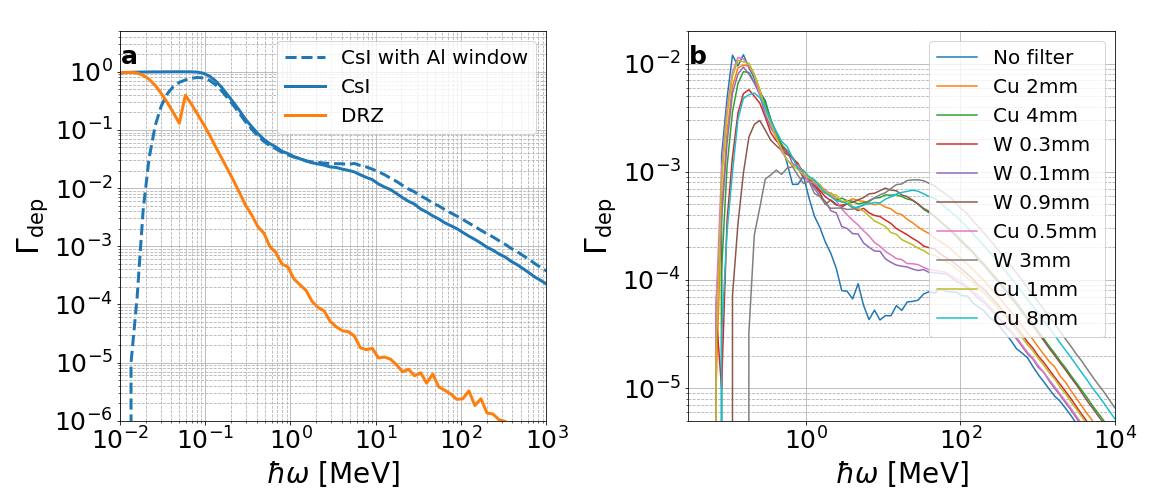}
    \caption{Spectral response of GAMMA1 scintillation screens (a) and of the GAMMA2 scintillation screen (DRZ-fine) with different conversion filters (b), as computed using GEANT4. See text for details regarding the filter and scintillator layout.}
    \label{fig1}
\end{figure}

Figure~\ref{fig1}(a) shows an example of the spectral response of two GAMMA1 scintillators as computed using GEANT4, with and without the effect of the Al exit window. For each photon energy, $10^6$ photons were used to compute the averaged energy deposited in the scintillation screen. We observe that at low energies ($10$-$20 \;\rm keV$) the Al window absorbs the photons and therefore no energy is deposited in the scintillator, meaning that those photons cannot be detected. For high energy photons ($\gtrsim 2 \;\rm MeV$) the Al window has the opposite effect on the amount of energy deposited in the scintillator: the secondary particles produced during the passage through the Al exit window reach the detectors and increase the amount of energy deposited in the scintillator. 

\subsection{GAMMA1}
The GAMMA1 detector at FACET-II is a scintillation-based X-ray and $\gamma$-ray detector designed to measure the integrated radiation yield and its angular distribution. This detector has two scintillation screens that can be individually inserted into the photon-axis: a DRZ-FINE$^{\rm TM}$ screen (manufactured by Mitsubishi Chemical Group) and a pixelated CsI array (manufactured by Epic-Crystal). The CsI array, formed by $165 \times 165$ crystals of $0.5 \times 0.5 \times 3\;\rm mm^{3}$ size, offers better sensitivity than the DRZ-FINE, both in terms of the spectral response (see Fig.~\ref{fig1}a) and light output (number of scintillation photons emitted for a given amount of deposited energy). This better sensitivity comes at the cost of a worse spatial resolution, given by the transverse size of an individual CsI crystal (0.5 mm). The visible light emitted by the scinitillator is imaged via a Nikon NIKKOR 50mm f/1.2 objective on an Allied Vision Manta G-125 GigE camera.

Similarly to the effect of the Al window on the spectral responses discussed above, a foil of high-Z material, such as 0.1 mm tungsten, can be installed on the upstream face of the DRZ-FINE scintillator in order to increase the sensitivity of the DRZ-fine to high photon energies ($\gtrsim 2$ MeV).

\subsection{GAMMA2}
\label{sec_G2}
The GAMMA2 detector at FACET-II is a scintillation-based X-ray and $\gamma$-ray detector designed to assess the spectral distribution of the incoming high-energy photons. It consists of a set of filters placed immediately prior to a second DRZ-FINE scintillating screen centered on the photon axis at $\sim 70$ cm distance downstream of the GAMMA1 scintillator (see Fig.~\ref{fig_intro}). The set of filters, distributed as in an axis-symmetric pie chart around the photon axis (see Fig.~\ref{fig5}(a)), are glued onto the upstream face of the GAMMA2 scintillating screen. The rear face of the GAMMA2 scintillating screen is imaged via a Nikon NIKKOR 50 mm f/1.2 objective on a Manta G-125 GigE camera.

The working principle of this detector relies on the modification of the spectral response by the different filter materials. For X-ray photons with energies below 100 keV, a pair of Ross filters~\cite{Ross:28} can be used to accurately measure the amount of radiation at a given spectral range. For higher photon energies, step filters of different materials and thicknesses have been used to reconstruct the incoming spectra from the transmission rates~\cite{Kneip_2010}. However, as explained earlier, for photon energies above 1 MeV these filters act as converters, and thus both the absorption and the secondary particle production need to be taken into account. The conversion phenomenon occurring at these high photon energies actually plays a central role to extend the GAMMA2 detector sensitivity up to the 10's of MeV energy range, where the absorption/transmission detectors are very insensitive. 

For the first set of user-assisted commissioning runs at FACET-II, a set of Cu and W step filters were mounted on the GAMMA2 detector. One of the filter placements, referred as "Gap" in Fig.~\ref{fig5}(a), was left empty for normalisation purposes. The associated spectral responses are shown on Fig.~\ref{fig1}(b). It should be noted that the Al window as well the GAMMA1 scintillation are included in the simulations performed to compute these spectral responses, which allowed for an accounting of their absorption and production of secondary particles. These curves show the trend explained above: for low incident photon energies ($\lesssim 1\;\rm MeV$) the thicker filters lead to a lower deposited energy in the scintillator, and thus lower signal (transmission mode), whereas for high incident photon energies ($\gtrsim 1\;\rm MeV$) the thicker filters lead to a stronger signal (conversion mode). As will be shown in the following sections, comparing the relative signals on the GAMMA2 scintillator behind each filter can be used, with the help of simulations, to assess the spectral distribution of the incoming X-ray and $\gamma$-ray photons. 

It should be noted that the choice of an axis-symmetric pie distribution of the GAMMA2 filters is optimised for cylindrically symmetric radiation profiles. Yet, deviations from this ideal  distribution can be corrected during the data analysis by weighting the signals after each of the GAMMA2 filters using the corresponding GAMMA1 angular distribution recorded upstream.

\section{Commissioning results}
\label{Bremss}
The first set of consistent measurements of high-energy photons at FACET-II was carried out by inserting Al foils into the electron beam axis at the IP to produce bremsstrahlung photons. These Al foils of thicknesses ranging from 0.1 mm to 2 mm have been installed at the IP of FACET-II for the ``Near-field-CTR-based self-focusing in beam-multifoil collisions" E332 experiment~\cite{2021_Archana}. This experiment aims to measure a significant production of $\gamma$-ray photons when the Near-Field-CTR effect dominates the beam-solid interaction over multiple scattering, leading to a strong focusing effect on the beam and a high transfer efficiency from the electron beam energy to the $\gamma$-ray radiation. 

\begin{figure}
    \centering
    \includegraphics[width=0.48\textwidth]{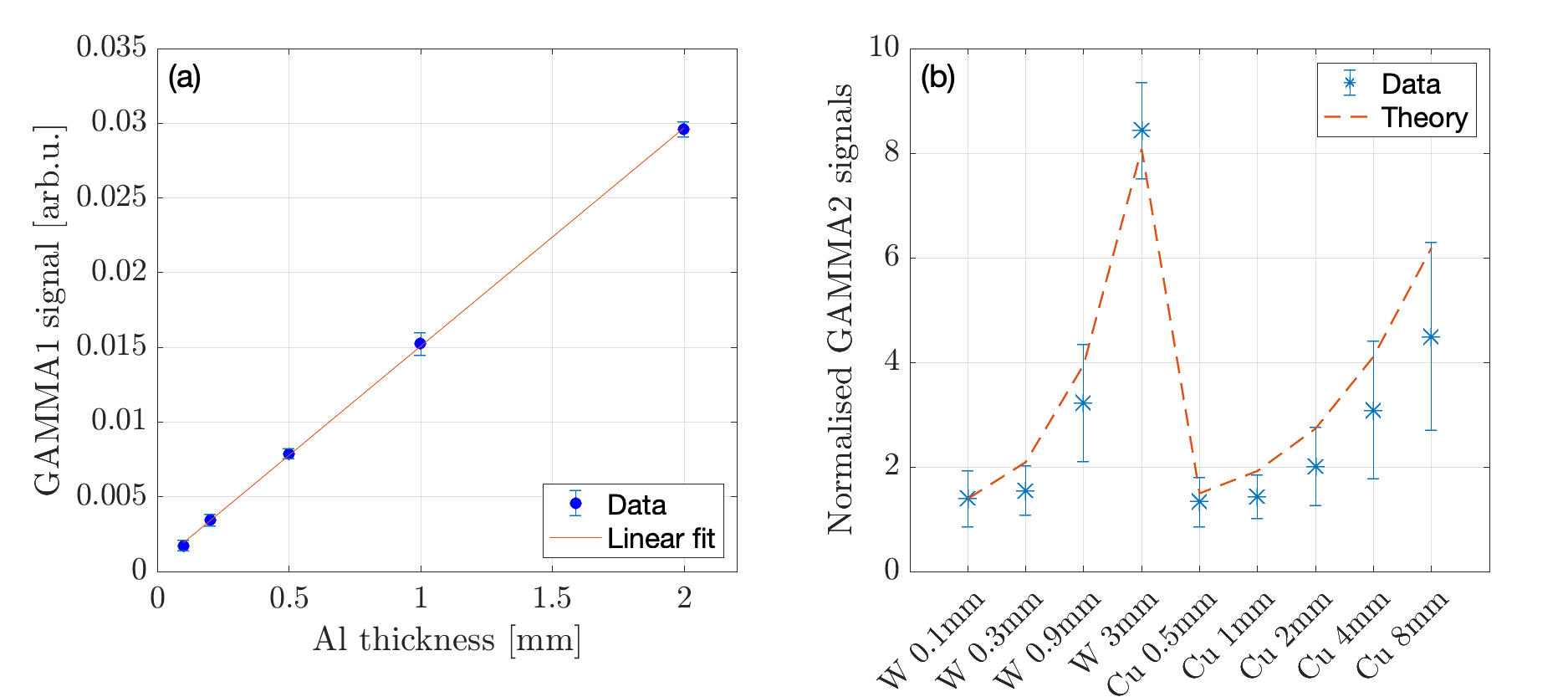}
    \caption{(a) Integrated GAMMA1 signal of Bremsstrahlung photons for different Al foil thicknesses (blue dots) and linear fit (red line) (b) Experimental and theoretical GAMMA2 signals of Bremsstrahlung photons produced with a W target of 1 mm thickness inserted at the IP.}
    \label{fig2}
\end{figure}

During the initial commissioning runs the beam parameters were such that the photons produced in the beam-solid interactions were originated predominantly via bremsstrahlung. By inserting Al foils of different thicknesses, the linearity of the detectors was assessed. The result of this test for GAMMA1 with the CsI scintillating screen is shown in Fig.~\ref{fig2}(a), indicating an acceptable linear relation between the GAMMA1 signal and the thickness of the Al foil used to produce bremsstrahlung radiation.

The GAMMA2 signal produced by these photons behind each filter was also recorded and compared with the theoretical predictions. For this analysis, a 1 mm thick W foil was inserted at the IP to maximize the flux of bremsstrahlung photons. The result of this analysis is shown in Fig.~\ref{fig2}(b). In this plot, both the simulated signals and the experimental signals are normalised by the corresponding GAMMA1 signals as explained in Sec.~\ref{sec_G2}, but also by the no-filter signal. For the theoretical values, the spectral responses shown in Fig.~\ref{fig1}(b) are applied to the analytical bremsstrahlung spectrum~\cite{Jackson:100964} produced by 10 GeV electrons propagating through a W foil. A fair agreement is observed between the experimental measurement and the theoretical values, benchmarking the modelling of the GAMMA2 detector described above. For this Bremsstrahlung measurements the GAMMA2 detector works in conversion mode, i.e. the thickest filters have the highest signals.

\section{Probing Strong-field QED at FACET-II}

Similarly to the seminal E144 experiment at the FFTB facility~\cite{Burke_PhysRevLett.79.1626}, FACET-II hosts an experimental campaign to study the Strong-Field regime of QED in the collision of a 10-TW class laser pulse with the 10-13 GeV electrons. 

\begin{figure}
    \centering
    \includegraphics[width=0.48\textwidth]{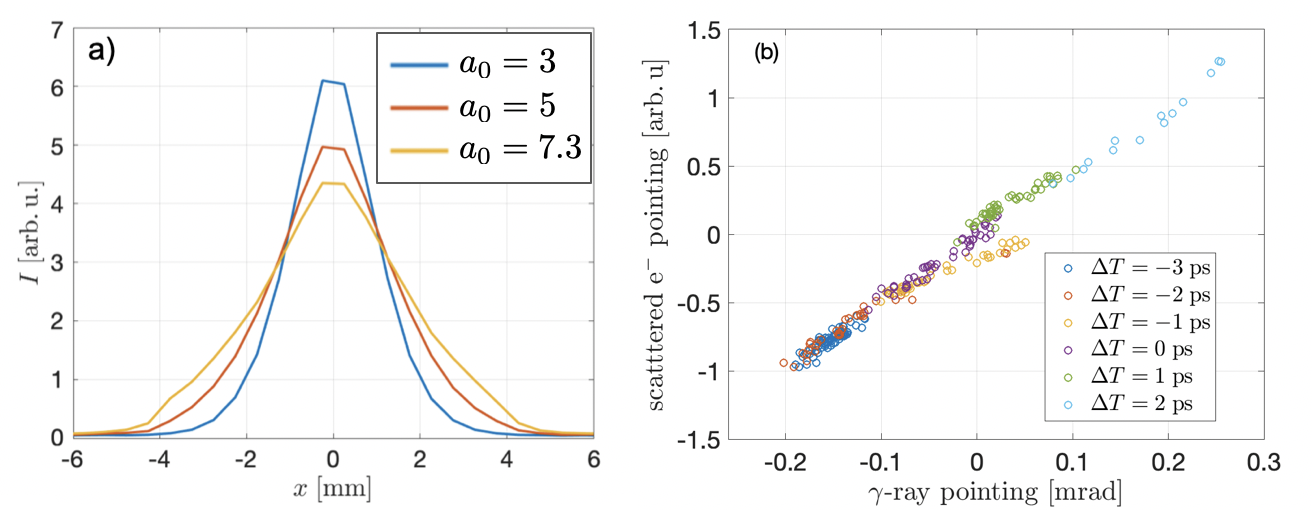}
    \caption{(a) Simulated horizontal distribution of the GAMMA1 signal of the high energy radiation produced during laser-beam collisions for three different values of laser normalised vector potential $a_0$. (b) Correlation between horizontal angle of laser-scattered electrons and $\gamma$-ray pointing of a temporal synchronisation scan of laser-beam collisions.}
    \label{fig4}
\end{figure}

The main goal of this campaign is to observe the transition from perturbative to non-perturbative electron laser interaction as well as to observe electron-positron pair production in the tunneling regime~\cite{Meuren_arxiv_2020,Meuren_PhysRevD_2016}. 
In these collisions, the beam electrons that are scattered by the electromagnetic field of the laser pulse emit high energy photons via (non)linear Compton scattering~\cite{Corde_RevModPhys.85.1,Bula_PhysRevLett_1996}. If detected, these photons can be used to retrieve information about the collision. For instance, theoretical calculations predict that when $a_0>1$ ($a_0$ being the normalised laser vector potential) the divergence of the inverse-Compton photons is proportional to $a_0$~\cite{Yan_NatPhot_2017}. Therefore measuring the divergence of the emitted photons can be used to track energy jitters of the laser and the true $a_0$ experienced by the colliding electrons on a shot-to-shot basis~\cite{Har-Shemesh_12_Optica}. It should nevertheless be noted that for the cases of few-cycle pulses or pulses significantly deviating from a Gaussian shape a more detailed analysis is needed, requiring a good experimental characterisation of the laser parameters. 
In order to test if the GAMMA1 detector is sensitive to this variation of $\gamma$-ray divergence for the FACET-II experimental configuration, GEANT4 simulations were performed to compute the spatially resolved energy deposition on the pixelised CsI crystal. In this simulation, the incoming photons were initialised using a Monte-Carlo algorithm that produced the angular and spatial distributions dictated by numerical QED simulations \cite{2017_Tamburini,2022_Monteflori_Arxiv}. Figure~\ref{fig4}(a) shows the horizontal (over the laser-polarisation direction) lineout of the simulated GAMMA1 signals for three different values of $a_0$. The root-mean-square values of this distribution confirms the linear relation between the measured $\gamma$-ray divergence and $a_0$.

During the commissioning runs on August 2022, first laser-electron collisions were achieved at the IP of FACET-II. Scattered electrons, with up to 2.5 GeV energy loss as well as the associated $\gamma$ photons were recorded by the different detectors of the experimental area. In this initial configuration, a laser pulse of $a_0\lesssim 1$ collided with the 10 GeV electron beam at a horizontal angle of $\approx 30^\circ$. 
At the collision point the electron-beam waist was much
larger than the $\approx 2\;\rm \mu m$ laser focal spot, and the electron beam waist was not set at the laser-IP. Preliminary analysis of these collisions suggests that  
 linear Compton scattering was dominant in the measured $\gamma$-ray signal. Yet, scans of the spatial and temporal laser-beam overlap were performed and allowed for characterisation of the collision parameters. In these scans, the GAMMA1 detector provided a clean signature of the collision. 

Figure~\ref{fig4}(b) shows data from a temporal overlap scan that manifests the correlation between the measured $\gamma$-ray horizontal pointing and the pointing (angle) of the scattered electrons as measured from the FACET-II electron spectrometer. It should be noted that, due to the $30^{\circ}$ angle of incidence, the temporal scan also encodes information on the horizontal spatial electron distribution. Both the $\gamma$-ray pointing and the scattered electron angle are measured in the non-dispersive plane of the electron spectrometer, which is along the laser polarisation direction. The scattered electron pointing was measured at around $9 \;\rm GeV$, the spectrometer being set to image the $8\;\rm GeV$ electrons from the IP to the screen. The clear correlation observed in Fig.~\ref{fig4}(b) between the $\gamma$-ray pointing and the angle of the scattered electrons corresponds to a converging electron beam with negative transverse position-momentum correlation. As the laser delay $\Delta T$ is scanned, the collision happens at different transverse positions of the converging electron beam, resulting in the observed time-dependent electron and $\gamma$ angles. This data shows, in a similar way to a laser-wire scanner~\cite{BOSCO2008162}, how the detection of the high-energy inverse-Compton photon produced in laser-electron collisions can provide a valuable signature of the interaction that can complement and improve the measurements of the scattered electrons, the latter suffering from high background levels of non-scattered electrons.

\section{Beam-driven Plasma Wakefield Accelerator}

The E300 experiment, ``Energy Doubling of Narrow Energy Spread Witness Bunch while Preserving Emittance with a High Pump-to-Witness Energy Transfer Efficiency in a Plasma Wakefield Accelerator", is the main beam-driven plasma wakefield accelerator experiment that will explore the current experimental challenges of this accelerator technology at the FACET-II facility~\cite{Joshi_2018}. Among these challenges, the preservation of the transverse quality of the accelerated beam, i.e. of the normalised emittance, is one of next milestones for the application of plasma acceleration technologies to high-energy particle colliders. To achieve this goal, controlling the matching of the beam betatron oscillations in and out of the plasma~\cite{Ariniello_PhysRevAccelBeams.22.041304} as well as mitigating the transverse Hosing instability~\cite{Huang_PhysRevLett.99.255001} is needed. Recently, simulations have shown that the betatron radiation produced by the accelerated beam can be used to diagnose the presence and mitigation of these two sources of emittance growth (mismatch propagation and Hosing instability)~\cite{2019_PTRS_SanMiguel}. The non-destructive nature of this novel diagnostic could help optimising the beam-plasma interaction and to preserve the emittance of the accelerated beam.

During the initial commissioning runs of this experiment, the region of the IP was filled with $\rm H_2$ gas with static pressures $P$ ranging from 0.05 to 5 Torr. Similarly to what was done at FACET-I~\cite{Corde_2016_NatComm}, the beam was able to field-ionise the $\rm H_2$ molecules over $\approx 3$ meters and drive strong plasma waves. Despite the large shot-to-shot fluctuations of the beam's longitudinal parameters due to the development of microbunching instabilities along the LINAC, significant energy loss and betatron radiation was consistently observed at every $\rm H_2$ pressure. Furthermore, multi-GeV particle acceleration was measured and is currently under study. 

\begin{figure}
    \centering
    \includegraphics[width=0.5\textwidth]{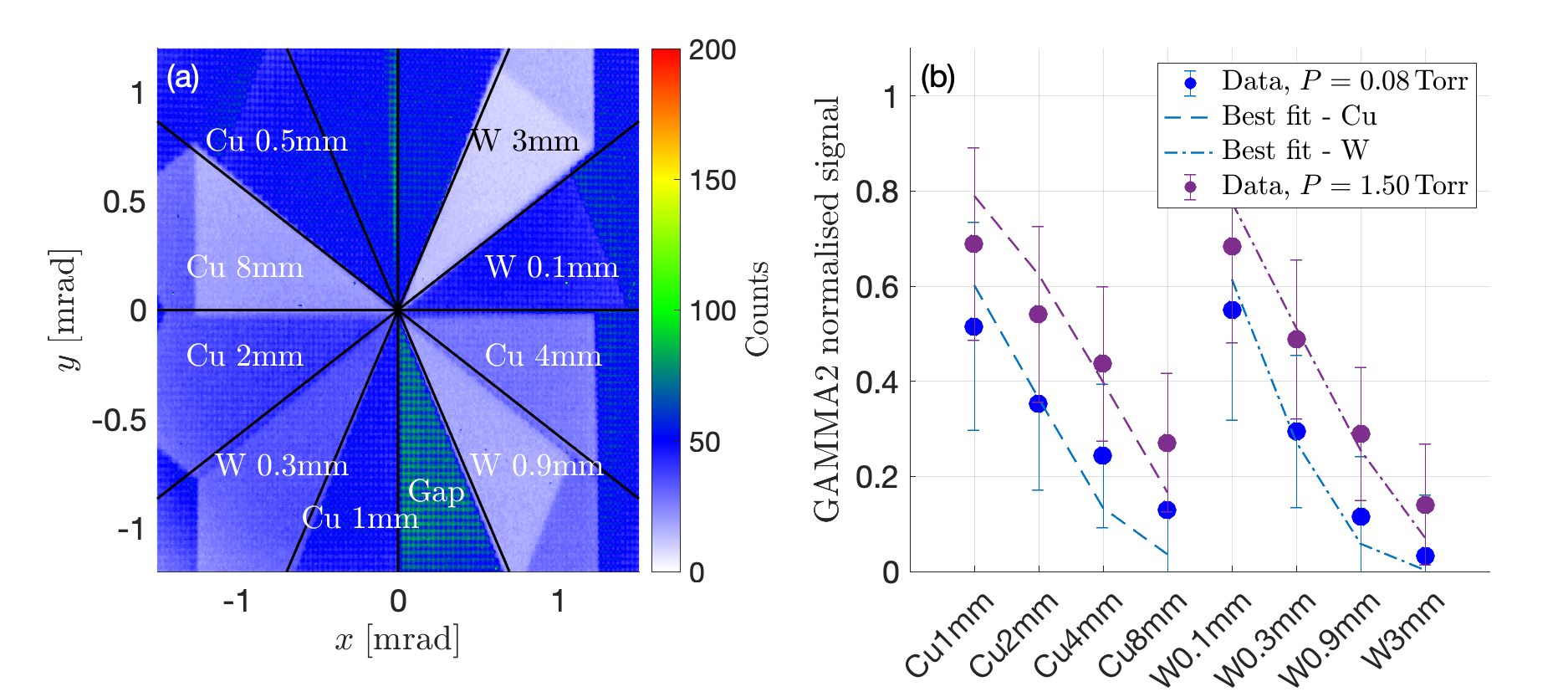}
    \caption{(a) GAMMA2 image of betatron radiation produced in a beam ionised $\rm H_2$ plasma of static pressure $P=1.5\;\rm Torr$, together with labels of the material and thickness of each filter. (b) Averaged GAMMA2 signals behind each filter at $P=0.08\;\rm Torr$ (blue dots) and $P=1.5\;\rm Torr$ (purple dots). The errorbars represent the standard deviations. The fitted values (see text) are plotted in dashed lines.}
    \label{fig5}
\end{figure}

From the GAMMA2 measurement of these betatron photons it was possible to retrieve a fitted critical frequency $\omega_c$ of a synchrotron-like spectrum. Results of this study are shown in Fig~\ref{fig5}. Figure~\ref{fig5}(a) shows a GAMMA2 image of the scintillation signal behind each filter. The observed hierarchy of these signals, i.e. higher signal after the thinner filters, indicates that photon absorption dominates over secondary particle production in the GAMMA2 filters. Therefore the detector works in transmission mode, in contrast to the conversion mode of the bremsstrahlung measurements [Fig.~\ref{fig2}(b)], and thus most of the betatron radiation photons are below the $\sim 1\;\rm MeV$ limit mentioned in Sec.~\ref{sec_G2}.

After averaging the signal behind each GAMMA2 filter over a selection of events (the 20 events with the highest $\gamma$-ray yield at each $\rm H_2$ pressure), the normalised signals - see Sec.~\ref{Bremss} for details of the normalisation procedure - are plotted in Fig.~\ref{fig5}(b) for $P=0.08$ Torr and $P=1.5$ Torr. Despite the large errorbars originating from the aforementioned shot-to-shot fluctuations, the comparison of the GAMMA2 signals for the two different pressures shows that, as pressure is increased, a much larger relative increase of the GAMMA2 signal behind the thicker filters (a factor of $\approx 4$ for W 3 mm) than after the thinner filters (a factor of $\approx 1.2$ for W 0.1mm) is observed. Qualitatively, this feature is explained by the shift of the betatron spectrum towards higher energies due to higher plasma densities at higher $\rm H_2$ pressure, which in turn results in more transmission of the higher energy photons and a more significant secondary particle production on the thicker filters.

A more quantitative analysis of this data is carried out by fitting a synchrotron spectrum, defined by a critical frequency $\omega_c$, to the experimental data~\cite{Ta-Phuoc:2012vb,Kneip_2010}. For each pressure, two fits are performed, one using the Cu filters and the other one using the W filters. For $P=0.08\;\rm Torr$ ($n_p\approx 2.5\times 10^{15}\;\rm cm^{-3}$ assuming single $\rm H_2$ ionisation), the fitted critical frequencies are $7\;\rm keV$ and $29\;\rm keV$ for the Cu filter and for the W filters respectively. The corresponding fitted signals are plotted with the blue dashed lines on Figure~\ref{fig5}(b). For $P=1.5\;\rm Torr$ ($n_p\approx 5\times 10^{16}\;\rm cm^{-3}$ assuming single $\rm H_2$ ionisation) the fitted critical frequencies are $24\; \rm keV$ and $123\;\rm keV$ respectively. It should be noted that at this plasma density, significant energy losses of more than 5 GeV were observed, which can substantially modify the shape of the emitted spectra and is not taken into account in this fitting analysis. Moreover, a discrepancy between in the absolute values of the retrieved critical frequencies from the Cu and W filters is observed and is currently under study.

This preliminary analysis shows a first study of the sensitivity of the GAMMA2 detector for the reconstruction of the betatron spectra produced at FACET-II. With the future optimisation of the beam performance at FACET-II, mainly in terms of shot-to-shot stability thanks to the mitigation of longitudinal instabilities, as well as foreseen improvements in the data analysis, the reconstruction of the betatron spectrum should constrain the beam parameter space and associated beam dynamics in the plasma.

\section{Conclusions}

In this article, we have reported on the design, commissioning and first measurements of X-ray and $\gamma$-ray detectors for the initial phase of the new accelerator facility FACET-II. As a results of a collaborative simulation campaign, two scintillation-based detectors, named GAMMA1 and GAMMA2, have been manufactured and installed to measure the broadband high-energy radiation produced at the Interaction Point of different experimental campaigns hosted at the facility. During the first user-assisted commissioning runs of FACET-II, bremsstrahlung, linear Compton scattering, and betatron high energy photons have been produced and measured using the GAMMA1 and GAMMA2 detectors, allowing several commissioning tests and experimental studies of their working principles. The results of the studies presented here show the potential uses of these detectors in the context of several experimental campaigns, from Strong-Field QED to beam-driven plasma wakefield acceleration experiments.

\printbibliography

\end{document}